\documentclass
{use}
\usepackage{graphicx}
\usepackage{txfonts}

\newcommand{\be}{\begin{equation}}
\newcommand{\ee}{\end{equation}}

\title[Testing cosmological models with galactic brightness
       profiles]{Testing cosmological models with the brightness
       profile of distant galaxies}
\author[I.\ Olivares-Salaverri and M.\ B.\ Ribeiro]
       {I.\ Olivares-Salaverri$^{1,2}$\thanks{E-mail: iosalaverri@gmail.com 
       (IOS); mbr@if.ufrj.br (MBR)} and Marcelo B.\
       Ribeiro$^{2}$\footnotemark[1]\thanks{Corresponding author}\\~\\
       $^{1}$Valongo Observatory, Universidade Federal do Rio de Janeiro,
       Rio de Janeiro, Brazil\\
       $^{2}$Physics Institute, Universidade Federal do Rio de Janeiro,
       Rio de Janeiro, Brazil}
\begin{document}
\date{}
\pagerange{\pageref{firstpage}--\pageref{lastpage}} \pubyear{2021}
\maketitle
\label{firstpage}
\begin{abstract}
The goal of this work is to use observed galaxy surface brightness
profiles at high redshifts to determine, among a few candidates, the
cosmological model best suited to interpret these observations.
Theoretical predictions of galactic surface brightness profiles are
compared to observational data in two cosmological models, $\Lambda$CDM
and Einstein-de Sitter, to calculate the evolutionary effects of
different spacetime geometries in these profiles in order to try to
find out if the available data is capable of indicating the cosmology
that most adequately represents actual galactic brightness profiles
observations. Starting from the connection between the angular diameter
distance and the galactic surface brightness as advanced by Ellis and
Perry (1979), we derived scaling relations using data from the Virgo
galactic cluster in order to obtain theoretical predictions of the
galactic surface brightness modeled by the S\'{e}rsic profile at
redshift values equal to a sample of galaxies in the range $1.5
\lesssim z \lesssim 2.3$ composed by a subset of Szomoru's et al.\
(2012) observations. We then calculated the difference between theory
and observation in order to determine the changes required in the
effective radius and effective surface brightness so that the observed
galaxies may evolve to have features similar to the Virgo cluster ones.
Our results show that within the data uncertainties of this particular
subset of galaxies it is not possible to distinguish which of the two
cosmological models used here predicts theoretical curves in better
agreement with the observed ones, that is, one cannot identify a clear
and detectable difference in galactic evolution incurred by the galaxies
of our sample when applying each cosmology. We also concluded that the
S\'{e}rsic index $n$ does not seem to play a significant effect in the
evolution of these galaxies. Further developments of the methodology
employed here to test cosmological models are also discussed.
\end{abstract}

\section{Introduction}\label{intro}

Observational cosmology attempts to understand the large-scale matter
distribution in the universe and its geometry by basically following
two different methodologies. The first, known as the
\textit{direct-manner}, or \textit{data-driven}, approach, seeks to
describe what is actually observed without addressing the question
of why we observe such an universe the way it is, whereas the
\textit{theory-based}, or \textit{model-based}, one interprets the
observations based on explanations that can produce the observed
patterns (Ellis 2006).

The theory-based approach consists of assuming a model based on
a spacetime geometry and then determines the values of the free
parameters by comparing the theoretical predictions with astronomical
observations of distant objects. Currently, the most accepted
cosmological model, the $\Lambda$CDM cosmology, is based on this
theory-based approach. It concludes that the universe is almost entirely
made up of dark matter and dark energy, whose compositions are presently
unknown.

The data-driven approach of observational cosmology claims that we are
in principle capable of determining the spacetime geometry on the null
cone by means of astrophysical observations, that is, using data
available on the past null cone (Kristian \& Sachs 1966; Ellis et al.\
1985). One specific study that follows this direct manner methodology
was advanced by Ellis \& Perry (1979), who developed a very detailed
discussion connecting the galactic brightness profiles with cosmological
models. Their aim was to determine the spacetime geometry of the universe
by measuring the \textit{angular diameter distance} $d_{\scriptscriptstyle A}$,
also known as area distance, of distant galaxies through their surface
brightness photometric data. Such a task was, however, made very
difficult due to lack of detailed knowledge about the structure and
evolution of galaxies. To this day this difficulty still remains. 

Here we propose a method for testing cosmological models partially
based on Ellis \& Perry (1979) methodology, but less ambitious than
theirs. Our approach differs from these authors in the sense that we
do not aim to determine the entire spacetime geometry from observations.
Our goal is to seek consistency between detailed astronomical
observations of quantities describing actual galactic structures as
compared to their predictions made by a specific cosmological model.
To be precise, we start by assuming a cosmological model and then
discuss the consistency between the model's predictions and actual
observations of the surface brightness of distant galaxies and other
related quantities. The idea is to obtain a glimpse of the structure
and evolution of galaxies by means of observations of our local
universe, identify observing parameters which, in principle, are
cosmological-model independent, that is, independent of the spacetime
geometry, and then assume that galactic scaling relations do not
change significantly, say within $3\sigma$, with the redshift. In
this way we could select distant galaxies that form a homogeneous
class of objects, defined as a set of similar galactic properties
which can be found at different galactic evolutionary stages, such
as morphology, so that one can compare objects at different redshift
values. This implies in assuming that the variations in the scaling
relations are related to the variations in the intrinsic structural
parameters of a previously selected homogeneous class of objects
(Ellis et al.\ 1984). By starting with a cosmological model we are
able to assess to what extent the assumed cosmology affects actual
galaxy evolution modeling carried out in extragalactic astrophysics
where some cosmological model, nowadays the $\Lambda$CDM, is
implicitly assumed. 

In this paper we address the first step in this approach of
cosmological model testing using actual galactic data, whose basic
methodology was briefly advanced elsewhere (Olivares-Salaverri \&
Ribeiro 2009, 2010). The purpose here is to verify if a sample of
galactic brightness profiles vary with their predicted theoretical
ones when one changes the cosmological model, that is, if surface
brightness profiles are affected by the spacetime geometry. We adopt
two distinct cosmologies, the $\Lambda$CDM and, for simplicity in
this initial approach, the Einstein-de Sitter (EdS) model. We then
calculate photometric scaling relations using the Virgo cluster
galaxies by means of the Kormendy et al.\ (2009) data and use our
theory to predict the galactic surface brightness at high redshift
values in the two cosmological models, assuming that these scaling
relations do not change with the redshift. We then compare these
predictions with a subsample of high redshift galactic surface
brightness data of Szomoru et al.\ (2012; from now on S12).   

Our results show that the observed high-redshift galactic brightness
profiles differ from the theoretically predicted ones obtained by
assuming that they follow scaling relations derived from the Virgo
cluster. Such a difference occurs when the theoretical results are
obtained by using both cosmological models studied here. Therefore,
for these galaxies to change their features to the ones found in the
Virgo cluster, an intrinsic evolution must take place. Such an
evolution is similar in both cosmological models. Consequently, our
results do not allow us to conclude which of the adopted cosmologies
produce more or less evolution or is more suitable to represent the
process in which high redshift galaxies develop into local galaxies
having scaling relations similar to the ones observed in the Virgo
galactic cluster, at least as far the chosen particular subset of
S12 galaxies is concerned. 

The outline of the paper is as follow. In \S\ref{section2} we discuss
cosmological distance measures and their connections to astrophysical
observables and \S\ref{section3} shows how the surface
brightness of cosmological sources are connected to those 
distance measures. We present in \S\ref{section4} the received surface
brightness using the profile due to S\'{e}rsic (1968). In \S\ref{section5}
we calculate two photometric scaling relations of the Virgo cluster so that
the next section (\S\ref{section6}) shows our comparison, using the two
cosmologies assumed here, between the prediction of the surface brightness
obtained by means of these scaling relations and the observations of
S12 high redshift galaxies. In \S\ref{section7} we calculate in both
cosmological models how galaxies whose redshift values are equal to the ones
in our chosen subsample of S12 observations would have to evolve to end
up with features similar to the galaxies in the Virgo cluster. Finally,
in \S\ref{section8} we summarize the results and present our conclusions.  

\section{Cosmological distances}\label{section2}

We start by considering that source and observer are at relative motion
to one another. From the point of view of the source, the light beams
that travel along future null geodesics define a solid angle
$\mathrm{d}\Omega_{\scriptscriptstyle G}$ with the origin at the source and
have a transverse section area $\mathrm{d}\sigma_{\scriptscriptstyle
G}$ at the observer (Ellis 1971; see also Fig.\ 2 of Ribeiro 2005 where
cosmological distances are also discussed in some detail). 

The specific radiative intensity $F_{em}$ is the emitted, or
intrinsic, radiation  \textit{measured at the source} in a unit
2-sphere $S_{unit}$ lying in the locally Euclidean space at rest with
the source and centered at it, also assumed to radiate locally with
spherical symmetry. It is related to the intrinsic source luminosity
$L$ by, 
\begin{equation}
 L = \int_{S_{unit}} F_{em} {\mathrm d}\Omega_{\scriptscriptstyle G}
   = 4\pi F_{em}.  
\end{equation}
Let us now define $F_{re}$ as the flux radiated by the source, but
\textit{measured by an observer} located at some future time $t_0$
relative to the source. This is, of course, the received flux
in the area $\mathrm{d}\sigma_{\scriptscriptstyle G}$ at
rest with the observer and implies a certain distance between source
and observer, distance which is geometrically defined along a null
curve in an expanding spacetime where both source and observer are
located. Thus, the source luminosity is given by, 
\begin{equation}
 L = \int_{S_{\! \! phys}} (1+z)^2 F_{re} {\mathrm d}\sigma_{\scriptscriptstyle G}, 
\end{equation}
where $z$ is the redshift and $S_{\! \! phys}$ is the physical surface
receiving the flux, e.g., a detector. The factor $(1+z)^2$ appears
here because both source and observer have their geometrical locus in
a curved and expanding spacetime (Ellis 1971). Now, it has long been
known that the \textit{area law} establishes that the source intrinsic
luminosity is independent from the observer (Ellis 1971). Therefore,
these two equations are equal, yielding,
\begin{equation}
L = \int_{S} F_{em} \; {\mathrm d}\Omega_{\scriptscriptstyle G} =
\int_{S} (1+z)^2 F_{re} \; {\mathrm d}\sigma_{\scriptscriptstyle G},
\end{equation}
\begin{equation}
 F_{em} \; {\mathrm d}\Omega_{\scriptscriptstyle G} = const =
(1+z)^2 F_{re} \; {\mathrm d}\sigma_{\scriptscriptstyle G}.
\label{e1}
\end{equation}
Considering the source's viewpoint, we may define the \textit{galaxy
area distance} $d_{\scriptscriptstyle G}$ as (Ellis 1971; see also
Fig.\ 2 of Ribeiro 2005),
\begin{equation}
 {\mathrm d}\sigma_{\scriptscriptstyle G} = {d_{\scriptscriptstyle G}}^2
 {\mathrm d}\Omega_{\scriptscriptstyle G}.
\end{equation}
Thus, equation (\ref{e1}) becomes,  
\begin{equation}
F_{re} = \frac{F_{em}}{{d_{\scriptscriptstyle G}}^2(1+z)^2}=
\frac{L}{4\pi}\frac{1}{{d_{\scriptscriptstyle G}}^2(1+z)^2}.
\label{e2}
\end{equation}
The factor $(1+z)^2$ can be understood as arising from \textit{(i)}
the energy loss of each photon due to the redshift $z$, and 
\textit{(ii)} the lower measured rate of incoming photons due to 
time dilation (Ellis 1971). Since the galaxy area distance
$d_{\scriptscriptstyle G}$ appearing in equation (\ref{e2}) cannot
be measured as ${\mathrm d}\Omega_{\scriptscriptstyle G}$ is defined
at the source, we need to change this equation into another one
containing measurable quantities. This can be done as follows.

{From} the point of view of the observer, the light beams that travel
along its past null geodesics leave the source and converge to the
observer, defining a solid angle $d\Omega_{\scriptscriptstyle A}$ with
the origin at the observer and having transverse section area
$d\sigma_{\scriptscriptstyle A}$ at the source (Ellis 1971; see also
Fig.\ 1 of Ribeiro 2005). Then we can define the \textit{angular diameter
distance} $d_{\scriptscriptstyle A}$ as being given by,
\begin{equation}
 {\mathrm d}\sigma_{\scriptscriptstyle A} = {d_{\scriptscriptstyle A}}^2
 {\mathrm d}\Omega_{\scriptscriptstyle A}.
\label{eGeometriaA}  
\end{equation}
Now we can use the \textit{reciprocity theorem}, due to Etherington 
(1933; see also Ellis 1971, 2007), to relate $d_{\scriptscriptstyle G}$
to $d_{\scriptscriptstyle A}$. This theorem is written as follows, 
\begin{equation}
{d_{\scriptscriptstyle G}}^2 = (1+z)^2 {d_{\scriptscriptstyle A}}^2.
\label{recip}
\end{equation}
Thus, it is now possible to connect the flux received by the
observer and the angular diameter distance by combining equations 
(\ref{e2}) and (\ref{recip}), yielding 
\begin{equation}
 F_{re} = \frac{F_{em}}{{d_{\scriptscriptstyle A}}^2(1+z)^4}.
\label{efluxosRE}
\end{equation}
The received flux $F_{re}$ and the redshift $z$ are astronomically
measurable quantities. So, if the angular diameter distance
$d_{\scriptscriptstyle A}$ is somehow determined astronomically, or
obtained from theory as a function of $z$, then the intrinsic flux
$F_{em}$ and, therefore, the intrinsic luminosity $L$ are both
determined for all redshifts.
 
\section{Connection with the surface photometry of cosmological sources}
\label{section3}

As discussed in \S \ref{section2}, equation (\ref{efluxosRE}) 
connects the received and emitted fluxes of sources located in a curved
spacetime, but that expression is valid for point like sources. Galaxies,
however, form extended sources of light and their characterization requires
defining another quantity, the surface brightness, better suited to describe
them. The \textit{received surface brightness} $B_{re}$ is defined as the
ratio between the received flux and the observed solid angle of the galaxy, 
\begin{equation}
 B_{re} \equiv \frac{F_{re}}{{\mathrm d}\Omega_{\scriptscriptstyle A}}. 
\end{equation}
Considering equations (\ref{eGeometriaA}) and (\ref{efluxosRE}), this
expression can be rewritten as, 
\begin{equation}
B_{re} = \frac{F_{em}}{{\mathrm d}\sigma_{\scriptscriptstyle A}}
\frac{1}{(1+z)^4}.
\label{def1}
\end{equation}
If we define the \textit{emitted surface brightness} $B_{em}$ as the
intrinsic flux of the source $F_{em}$ per area unit in the rest frame
of the source, we have that 
\begin{equation}
 B_{em} \equiv \frac{F_{em}}{{\mathrm d}\sigma{\scriptscriptstyle A}}.
\end{equation}
Thus, equation (\ref{def1}) in fact connects the received and emitted 
surface brightness, as follows,  
\begin{equation}
B_{re}=\frac{B_{em}}{(1+z)^4}.
\label{erelaBrilho}
\end{equation}
This expression is simply the \textit{Tolman surface brightness test}
for cosmological sources, showing that galactic surface brightness does
not depend on the distance. This equation also shows that if there is
no significant cosmological effects, that is, if source and observer are
close enough to be considered at rest with one another and Newtonian
approximation is valid, then the source redshift is not significant. In
this case there is no cosmological contribution ($z \sim 0$) and
$B_{re}=B_{em}$. This can be considered as a consequence of the Liouville
theorem (Bradt 2004). It is also worth mentioning that several authors
name the radiation measured by the observer as \textit{intensity} $I$,
and the radiation emitted by the source as \textit{surface brightness}
$B$, instead of terms adopted here, respectively, received surface
brightness $B_{re}$ and emitted surface brightness $B_{em}$. This is
often the case in texts where General Relativity is not considered.

Actual astronomical observations are carried out in observational
bandwidths and, therefore, the equations discussed so far should take
this fact into account. The \textit{specific received surface brightness}
$B_{re,\nu_{re}}$ gives the amount of radiation received by the observer
per unit solid angle measured at the observer in the frequency range
$\nu_{re}$ and $\nu_{re}+{\mathrm d}\nu_{re}$. Clearly $B_{re}=\int_0^\infty
B_{re,\nu_{re}} {\mathrm d} \nu_{re}$. 
Considering equation (\ref{erelaBrilho}), we have that,
\begin{equation}
B_{re,\nu_{re}} {\mathrm d}\nu_{re} = \frac{B_{em,\nu_{em}}}{(1+z)^4}
 \, {\mathrm d}\nu_{em},
\label{def2}
\end{equation}
where we had defined the \textit{specific emitted surface brightness}
as follows,
\begin{equation}
B_{em,\nu_{em}}=B_{em} J(\nu_{em}).
\label{def3}
\end{equation}
Here $J(\nu)$ is the galactic \textit{spectral energy distribution} (SED)
giving the proportion of radiation at each frequency, being normalized
by the condition
\begin{equation}
\int_0^\infty J(\nu)\,{\mathrm d}\nu=1.
\label{def4}
\end{equation}
From our definitions it also follows that $B_{em}=\int_0^\infty
B_{em,\nu_{em}} {\mathrm d} \nu_{em}$. 
We need now to relate the received and emitted frequencies. This is
accomplished by the definition of the redshift,
\begin{equation}
\nu_{em}=\nu_{re}(1+z), 
\label{def5}
\end{equation}
implying that the SED of the source is observed according to
$J(\nu_{em})= J\left[\nu_{re}(1+z)\right]$
and equation (\ref{def2}) can be rewritten as follows,
\begin{equation}
B_{re,\nu_{re}} {\mathrm d}\nu_{re} = \frac{B_{em,\nu_{em}}}{(1+z)^3} \,
{\mathrm d}\nu_{re}.
\label{eRelaB}
\end{equation}

The variables in the equation above depend on some implicit parameters.
In order to reveal these dependencies, let us start by assuming our
galaxy as having spherical symmetry with space points defined by the
radius $R$. Furthermore, if this galaxy has a circular projection in the
celestial sphere, any angle $\alpha$ \textit{measured by the observer}
corresponds to the radius $R$ \textit{in the source} by means of the
angular diameter distance at a given redshift. Hence, we have that 
(see Ellis \& Perry 1979, Fig.\ 1),
\begin{equation}
 R(z)=d_{\scriptscriptstyle A}(z) \; \alpha.
\label{eRadiusDist}
\end{equation}

This expression is in fact a simplification of Eq.\ (\ref{eGeometriaA})
where area and solid angle are respectively approximated to length and
angle so that $d_{\scriptscriptstyle A}$ can be estimated observationally
(Ellis 1971, Ribeiro 2005). Indeed, it is used in observational cosmology
tests under the name ``angular diameter redshift relation'' since the
angular diameter distance has all cosmological information. So, different
values of the angular diameter distance are related to different
cosmological models. As mentioned above, in this paper the two chosen
cosmological models are EdS and $\Lambda$CDM.

The EdS cosmology has zero curvature and no cosmological
constant so, in this model the angular diameter distance may be written
as below,
\begin{equation}
 d_{\scriptscriptstyle A, EdS}(z)= 
\frac{2c}{H_0(1+z)}\biggl[1-\frac{1}{\sqrt{(1+z)}}
\biggr],
\label{distanciaporareaEdS}
\end{equation}
where $c$ is the light speed and $H_0$ is the Hubble constant.
This expression shows that $d_{\scriptscriptstyle A,EdS}$ reaches a
maximum value at $z=1.25$ and then starts decreasing, asymptotically
vanishing at the big bang singularity hypersurface.

In the case of the $ \Lambda$CDM model, several tests have been
carried out in the last few years to measure the parameter values of the
model leading to such a degree of accuracy that it became known as the
\textit{concordance model}, being the most accepted cosmology nowadays.
Those tests involve studies of the cosmic microwave background radiation,
baryonic acoustic oscillations and type Ia supernovae. Komatsu et al.\
(2009) presented values for several parameters in this cosmology to a
high degree of accuracy, such as $H_0=71.8$~km~s$^{-1}$~Mpc$^{-1}$,
$\Omega_{m_0}=0.273$ and $\Omega_{\Lambda}=0.727$. We used these values
to calculate numerically the angular diameter distance in both cosmologies,
that is, $d_{\scriptscriptstyle A,\Lambda CDM}$ and $d_{\scriptscriptstyle
A,EdS}$, where the latter is evaluated directly from equation
(\ref{distanciaporareaEdS}).

Returning to equation (\ref{eRelaB}), other implicit parameter
dependence also occurs in both specific surface brightness. The
received one depends on the \textit{observed} parameters $\alpha$, $z$
and $\nu_{re}$, so that we should write it as $B_{re,\nu_{re}}=
B_{re,\nu_{re}}(\alpha, z)$. The specific emitted surface brightness
depends on the \textit{source} parameters $R$, $\nu_{em}$ and, if allowed
for the intrinsic evolution of the source, also in the $z$. Thus,
$B_{em,\nu_{em}} = B_{em}(R,z) \, J[\nu_{re}(1+z),R,z]$. With these
dependencies, equation (\ref{eRelaB}) turns out to be written as the
expression below (Ellis \& Perry 1979),
 \begin{equation}
  B_{re,\nu_{re}}(\alpha,z) 
= \frac{B_{em}(R,z)}{(1+z)^3} \, J \left[\nu_{re}(1+z), R, z\right].
\label{eRelaBDEF}
\end{equation}
Note that this equation is completely general, i.e., valid for
\textit{any} cosmological model. 

Next we shall show how the surface brightness $B_{em}(R,z)$ can
be characterized by means of the S\'{e}rsic (1968) profile and obtain
an explicit expression for the received surface brightness.

\section{Received S\'{e}rsic surface brightness}\label{section4}

The S\'{e}rsic profile was not commonly considered among astronomers after
its proposal. Gradually, however, some authors started to claim that the
S\'{e}rsic index is not simply a parameter capable of providing a better
mathematical fit, but that it does have a physical meaning (Ciotti 1991;
Caon et al.\ 1993; D'Onofrio et al.\ 1994). Nowadays, this profile
seems to be more accepted since one can find in the recent literature
several papers using it as well as relating its parameters to other
astrophysical quantities (Davies et al.\ 1988; Prugniel \& Simien 1997;
Ciotti \& Bertin 1999; Trujillo et al.\ 2001; Mazure \& Capelato 2002;
Graham 2001, 2002; Graham \& Driver 2005; La Barbera et al.\ 2005;
Coppola et al.\ 2009; Chakrabarty \& Jackson 2009; Laurikainen et al.\
2010). In view of this, we believe that this profile is the most suitable
for the purposes of this paper. 

The S\'{e}rsic profile can be presented in two slightly different
parametric formats, although both of them characterize the same
brightness profile. The difference lies in the interpretation of
the parameters. The first one can be written as,
\begin{equation}
B_{\mathrm{S_1},em}(R,z) = B_{0}(z)\,\exp \left\{-
\left[\frac{R(z)}{a(z)} \right]^{1/n}\right\},
\label{eqSersic1}
\end{equation}
where $B_{0}$ is the brightness amplitude, $a$ is the scalar radius
and $n$ is the S\'{e}rsic index. Ellis \& Perry (1979) implicitly used
this form. The second way of writing the S\'ersic profile is given by
the following expression,
\begin{equation}
B_{\mathrm{S_2},em}(R,z)= B_{eff}(z)\,\exp \left\{ \displaystyle -b_{n}
\left[\left(\frac{R(z)}{R_{eff}(z)} \right)^{1/n} - 1 \right] \right\},
\label{eSersicprofile}
\end{equation}
{where $R_{eff}(z)$ is the effective radius, $B_{eff}(z)$ is the
brightness at the effective radius and $b_n$ is a parameter dependent
on the value of $n$.

The main difference between these two equations
is the nature of their parameters. In equation (\ref{eqSersic1}) $B_{0}$
and $a$ do not have a clear physical meaning, whereas the parameters
$B_{\scriptscriptstyle{eff}}$ and $R_{\scriptscriptstyle{eff}}$ 
appearing in equation (\ref{eSersicprofile}) are more easily interpreted.
$R_{\scriptscriptstyle{eff}}$ is defined as the isophote that contains
half of the total luminosity and $B_{\scriptscriptstyle{eff}}$ is the
value of the brightness in that radius (Ciotti 1991; Caon et al.\
1993). For this reason we believe that the second form above is more
appropriate for our analysis and we shall use it from now on.}

A relationship between the different parameters in equations 
(\ref{eqSersic1}) and (\ref{eSersicprofile}) can be obtained by
equating these two expressions, yielding,
\begin{equation}
 B_{0}(z) = B_{\scriptscriptstyle eff}(z) \, e^{b_n},
\label{eqConverBrilho}
\end{equation}
 \begin{equation}
a(z) = \frac{R_{\scriptscriptstyle eff}(z)}{{b_n}^n}.
\label{eqConverRad}
\end{equation}

{The evolution of the galactic structure is implicit in this profile
in view of the fact that both $B_{eff}(z)$ and $R_{eff}(z)$ are
redshift dependent. In addition, the connection of this profile to a
cosmological model and, therefore, to the underlying curved spacetime
geometry and its evolution occurs in the intrinsic radius $R(z)$ by
means of equation (\ref{eRadiusDist}). Thus, galaxies can possibly
experience two simultaneous evolutionary effects, intrinsic source
evolution and cosmological, or geometrical, evolution. Since at our
current knowledge of galactic structure both effects cannot be easily
separated, if they can be separated at all, from now on we shall assume
that $R(z)$ depends only on the underlying spacetime geometry given by
a chosen cosmological model. Furthermore, at first we shall not consider
a possible intrinsic evolution of the S\'{e}rsic index in the form
$n=n(z)$, because we assume that its change is produced via galactic
merger processes (Naab \& Trujillo 2006). So, the way that the
evolutionary dependency is defined in the emitted surface brightness
(Eq.\ \ref{eSersicprofile}) means that we are implicitly considering
galaxies belonging to a set with similar properties, or a homogeneous
class of objects. Departures from this class occur by smooth dependency
in the evolution of the intrinsic parameters such as $B_{eff}(z)$ and
$R_{eff}(z)$ (Ellis et al.\ 1984).}

{Let us now return to the properties of the S\'{e}rsic profile. 
There are analytical and exact expressions relating to $b_n$ and $n$.
Analytical expressions for $b_n$ were given by several authors
(Capaccioli 1989; Ciotti 1991; Prugniel \& Simien 1997), whereas
others worked out exact values for this parameter (Ciotti 1991;
Graham \& Driver 2005; Mazure \& Capelato 2002). Ciotti \& Bertin
(1999) analyzed the exact value form and concluded that a fourth
order expansion is enough to obtain good results, which are even
better than the values obtained from the analytical expressions.
Such an expansion is enough for the purposes of this paper and may
be written as below,} 
\begin{equation}
 b_n = 2n -\frac{1}{3} + \frac{4}{405n} + \frac{46}{25515n^2}.
\label{ebn4}
\end{equation}

Having expressed $B_{em}$ in terms of the S\'{e}rsic profile, 
we can now obtain the equation for the received surface brightness
$B_{re}(\alpha, z)$ since equation (\ref{eRelaBDEF}) gives the
relationship between the emitted and received surface brightness.
Considering the emitted brightness as modeled by the second form of
the S\'ersic profile (Eq.\ \ref{eSersicprofile}), we then substitute
the latter equation into the former and obtain the following expression,
\begin{eqnarray} 
B_{re,\nu_{re}}(\alpha,z) & = & \frac{B_{eff}(z)}{(1+z)^3} \;
J[\nu_{re}(1+z),R,z] \times \nonumber \\
 && \times  \exp{ \left\{ -b_{n} \left[
 {\left( \frac{R(z)}{R_{eff}(z)}\right)}^{1/n} - 1 \right] \right\} }.
\label{def10}
\end{eqnarray} 
Let us now define two auxiliary quantities required to study this
problem from an extragalactic point of view (Caon et al.\ 1993;
Graham \& Driver 2005),
\begin{equation}
\mu_{re,\nu_{re}} (\alpha,z) \equiv -2.5 \log \left( B_{re,\nu_{re}} \right),
\label{def8}
\end{equation}
\begin{equation}
\mu_{eff}(z) \equiv -2.5\log \left[ B_{eff}(z)\right].
\label{def9}
\end{equation}
Equation (\ref{def8}) is given in units of $[\mu_{re, \nu_{re}}]$ =
[mag/arc sec$^2$]. Considering these definitions, equation (\ref{def10})
can be rewritten as follows,
\begin{eqnarray}
\mu_{re, \nu_{re}}(\alpha, z) &=&  
\mu_{eff}(z)  +7.5\log(1+z) \nonumber \\ 
&& -2.5 \log \left\{ J\left[\nu_{re}(1+z), R, z\right]\right\} \nonumber \\ 
&& + \left[ \frac{2.5}{\ln(10)} \right] b_n\left \{\left[
\frac{R(z)}{R_{eff}(z)}\right]^{1/n} - 1 \right \}.
\label{eq.mhuTH}
\end{eqnarray}
This expression allows us to predict the galactic surface brightness in
a given redshift and in a specific cosmological model. Nevertheless, in
order to relate this expression with actual observations, we still need
further assumptions regarding galactic structure. Next we shall make use
of Kormendy et al.\ (2009) data to derive photometric scaling relations
of the Virgo galaxy cluster and suppose that these relations do not
change drastically out to the redshift range under study here.

\section{Photometric scaling relations of the Virgo Cluster}\label{section5}

In order to compare theoretical predictions of the surface brightness
profiles with observational data we require information regarding the
galactic structure, information which can be obtained after investigating
scaling relations that different kinds of galaxies follow in the local
universe. The most well-known of those relations are the ones that relate
luminosity with velocity dispersion, such as the \textit{Faber-Jackson
relation} (Faber \& Jackson 1976) for elliptical galaxies, and the
\textit{Tully-Fisher relation} (Tully \& Fisher 1977) for spiral galaxies.
Another, more general, scaling relation is \textit{the fundamental
plane} (Djorgovski \& Davis 1987) which correlates velocity dispersion
with effective brightness and effective radius instead of the luminosity.
It is worth mentioning that the fundamental plane is a generalization of
the \textit{Kormendy relation} (Kormendy 1977).

The parameters involved in these scaling relations are measured
through spectroscopic and photometric techniques. After knowing that the
S\'{e}rsic index and the velocity dispersion correlate, a fundamental
plane that only uses photometric parameters was proposed by Graham
(2002). This variant, called \textit{photometric plane}, uses parameters
which appear in the definition of the S\'{e}rsic  profile, i.e., the
S\'{e}rsic index $n$, the effective surface brightness $\mu_{eff}$ and
the effective radius $R_{eff}$. Since the study presented in this paper
deals with photometric parameters, we shall use the photometric plane to
obtain the galactic scaling relations required in our analysis.
Specifically, our focus will be on the galaxies belonging to the Virgo
cluster as their data (Kormendy et al.\ 2009) form the most exhaustively
studied galactic surface brightness dataset in the local Universe.

\subsection{Virgo cluster data}  

Kormendy et al.\ (2009) studied 42 galaxies of the Virgo cluster
on the V band, $\lambda_{V,eff}=(5450 \pm 880)$ \AA, whose morphological
types are elliptic (E), lenticular (S0) and spheroid (Sph). They used
the S\'{e}rsic profile to fit the observed surface brightness and, in the
authors' words, ``the S\'{e}rsic functions fit the main parts of 
the profiles of both elliptical and spheroidal galaxies astonishingly well
on large ranges in surface brightness.''

\begin{figure*}
\includegraphics[width=8.8cm]{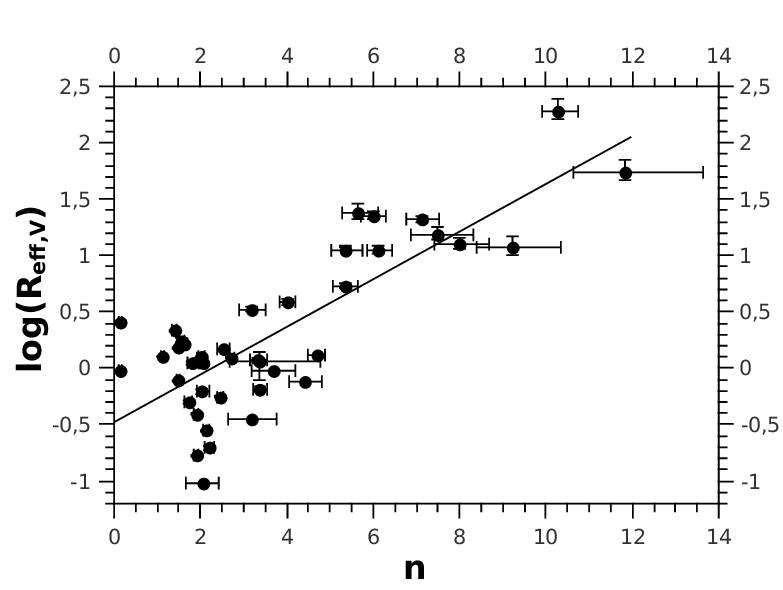}
\includegraphics[width=8.8cm]{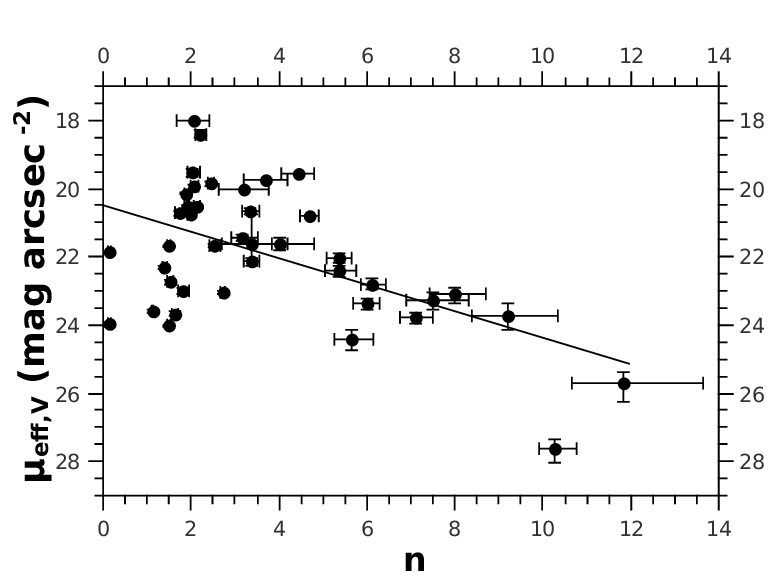}
\caption{\textit{Left}: Plot of the logarithm of the effective radius
$R_{eff,V}$ vs.\ the S\'{e}rsic index $n$ of galaxies from the
Virgo cluster. The straight line represents the linear fit whose values
are given in eq.\ (\ref{eq.KormendyReffN}). \textit{Right}: Plot of the
effective brightness $\mu_{eff,V}$ and the S\'{e}rsic index $n$ of
galaxies belonging to the Virgo cluster. The straight line represents
the linear fit whose values are given in eq.\ (\ref{eq.KormendyMhuN}).
\textit{Both}: Data taken from Kormendy et al.\ (2009).}  
\label{figuraKormendy}
\end{figure*}

As the observed data of the Virgo cluster is well fitted by the
S\'{e}rsic profile, the parameters obtained through these fittings can
be considered as reliable. Thus, using the parameters involved in the
photometric plane, $n$, $\mu_{eff}$ and $R_{eff}$, we carried out linear
fittings relating $\mu_{eff}$ to $n$ and $R_{eff}$ to $n$. Plots showing
these fittings are presented in Fig.\ \ref{figuraKormendy} and the fitted
parameters are as follows, 
\begin{equation}
\mu_{eff,V} = (0.38 \pm 0.09)n + (20.5 \pm 0.5),
\label{eq.KormendyMhuN}
\end{equation}
\begin{equation}
\log R_{eff,V} = (0.21 \pm 0.03)n - (0.5 \pm 0.1).
\label{eq.KormendyReffN}
\end{equation} 
The index $\scriptstyle V$ stands for the Virgo cluster.

Analyzing the plots we can observe that most galaxies with
low S\'{e}rsic index do not correlate. These galaxies are mostly of
spheroidal types and are the main contributors to the large errors
in the Y-axis. As our aim is to compare the prediction of the 
surface brightness with high-redshift observational data, if we knew
the morphological types of these high-redshift galaxies we could select
specific galaxy types to obtain the scaling relation. But, this is
usually not possible because galactic morphology at high redshifts is 
not as well established as in the local universe. Considering such
constraint, we resorted on using galaxies of the Virgo cluster whose
scaling relations were just derived in order to predict the surface
brightness at any redshift and compare it with high redshift galaxy
observations of S12. On top of the scaling relations, this is done by
employing the two cosmological models adopted here.

\section{High-redshift vs.\ predicted galactic surface brightness
         profiles in $\Lambda$CDM and EdS cosmologies}\label{section6}

In order to analyze the possible effects that different 
cosmological models can have in the estimated galactic evolution, we
shall compare theoretical predictions of the surface brightness data
with high-redshift galactic brightness profiles. An important issue
concerning high redshift galactic surface brightness is the depth
in radius of these images. Observing the universe at high redshift
involves lower resolution images, so obtaining a good sample of the
brightness profiles as complete as possible in radius is not an easy
task.

\subsection{The high redshift galactic data of Szomoru et al.\ (2012)}  

S12 were mainly interested in investigating if the interpretation of
the compactness of high redshift galaxies was due to a lack of deep
images in radius which would lead to a misinterpretation of the
compactness pattern. They observed a stellar mass limited sample of
21 quiescent galaxies in the redshift range $1.5 < z < 2.5$ having
${\cal{M}}_{*} > 5.10^{10}{\cal{M}}_{\odot}$ and obtained their surface
brightness using the S\'{e}rsic profile with high values in radius. We
chose S12 data for our purposes mainly because of these features.

S12 used NIR data taken with HST WFC3 as part of the CANDELS
survey. Specifically, the surface brightness profiles were obtained in
the $H_{160}$ band, $\lambda_{H,eff}=(13923 \pm 3840)$~\AA, that is,
comparable with the $V$ band rest-frame in the redshift range under
consideration. From the 21 galaxies we have selected a subsample having
S\'{e}rsic indexes grouped in three sets: $n \sim 1$, $n \sim 4$ and
$n \sim 5$. The last two sets are more numerous and are located in
several redshift values, especially the group with $n \sim 5$. Table
\ref{table1} shows our subsample with the identification number in the
S12 catalog and their respective S\'{e}rsic indexes and the redshifts.
\begin{table}
\caption[adasdq]
{Identification (ID) number, S\'{e}rsic index $n$ and redshift of the
galaxy subsample selected from S12 and used in this paper.} 
\begin{center}
\begin{tabular}{@{\extracolsep{\fill}}l c  r @{\extracolsep{\fill}}}
\hline \hline ID & $n$ & $z$ \\ \hline
2.856 & 1.20 $\pm$ 0.08 & 1.759 \\
3.548 & 3.75 $\pm$ 0.48 & 1.500 \\
2.531 & 4.08 $\pm$ 0.30 & 1.598 \\
3.242 & 4.17 $\pm$ 0.45 & 1.910 \\
3.829 & 4.24 $\pm$ 1.15 & 1.924 \\
1.971 & 5.07 $\pm$ 0.31 & 1.608 \\
3.119 & 5.09 $\pm$ 0.60 & 2.349 \\
6.097 & 5.26 $\pm$ 0.56 & 1.903 \\
1.088 & 5.50 $\pm$ 0.67 & 1.752 \\
\label{table1}
\end{tabular}
\end{center}
\end{table}

\subsection{Theoretical predictions of the surface brightness}

As discussed above, equation (\ref{eq.mhuTH}) allows us to
calculate the surface brightness theoretical prediction of a
hypothetical galaxy in a given redshift. We have already calculated
scaling relations for local galaxies which we assume to be maintained 
out to the observed S12 redshift values. However, to see if the
observed high redshift galaxies behave like the Virgo cluster ones,
i.e., if they obey the assumed scaling relations, we must calculate
the errors of the theoretical predictions in Eq.\ (\ref{eq.mhuTH}).

Errors for $\Delta \mu_{re,\nu_{re}}$ are estimated quadratically
as follows,
\begin{eqnarray}
\Delta \mu_{re, \nu_{re}} &=&  
\biggl [ \biggl (\frac{\partial \mu_{re, \nu_{re}}}{\partial \mu_{eff}}\Delta 
\mu_{eff}\biggr )^2 
 + \biggr (\frac{\partial \mu_{re, \nu_{re}}}{\partial z}\Delta z\biggr )^2 
+ \biggr (\frac{\partial \mu_{re, \nu_{re}}}{\partial n}\Delta n\biggr )^2 
\nonumber \\ 
&& + \biggr (\frac{\partial \mu_{re, \nu_{re}}}{\partial R}\Delta R\biggr )^2 
+ \biggr (\frac{\partial \mu_{re, \nu_{re}}}{\partial \log (R_{eff})}\Delta 
\log (R_{eff})\biggr )^2 \biggl ]^{1/2}.
\label{erromu}
\end{eqnarray}
Let us analyze individually each of the uncertainty terms in this
expression.

\subsubsection{$\Delta \mu_{eff}$}

The effective surface brightness is calculated from the 
linear fit of the Virgo cluster galaxy scaling relation, 
\begin{equation}
 \mu_{eff} = A_{\mu_{eff}}n +B_{\mu_{eff}},
\end{equation}
where $A$ and $B$ are the linear fit parameters given in
equation (\ref{eq.KormendyMhuN}). Its uncertainty yields,
\begin{equation}
 \Delta \mu_{eff} =\biggl [ \biggl( 
\frac{\partial \mu_{eff}}{\partial A_{\mu_{eff}}}\Delta A_{\mu_{eff}} 
\biggr )^{2} 
+ \biggl( 
\frac{\partial \mu_{eff}}{\partial n}\Delta n \biggr )^{2}
+\biggl( 
\frac{\partial \mu_{eff}}{\partial B_{\mu_{eff}}}
\Delta B_{\mu_{eff}} \biggr )^{2}
\biggr ]^{1/2}.
\end{equation}
This expression can can be rewritten as below,
\begin{equation}
\Delta \mu_{eff} = 
\biggl [ (n\Delta A_{\mu_{eff}})^2 + (A_{\mu_{eff}}\Delta n)^2 + 
(\Delta B_{\mu_{eff}})^2 \biggr ]^{1/2},  
\end{equation}
where $\Delta A$ and $\Delta B$ come from the linear
fit and $\Delta n$ comes from the observation.\\ 

\subsubsection{$\Delta z$}

Redshift uncertainty was given by S12 only if $z$ was measured
photometrically. However, some of S12's galaxies had their redshifts
measured spectroscopically, then the errors were not made
available. Since there is a mixture of photometric and spectroscopic
redshifts in S12's sample, we decided to avoid including them in our
calculations and have effectively assumed $\Delta z \sim 0$.

\subsubsection{$\Delta n$}

The value of S\'{e}rsic index and its error came directly from
the observations shown in S12 (see Table \ref{table1}). 

\subsubsection{$\Delta R$}

The projected radius is obtained by means of equation
(\ref{eRadiusDist}), whose quadratic uncertainty may be written as
follows, 
\begin{equation}
\Delta R = \biggl [ \biggl( 
\frac{\partial R}{\partial \alpha}\Delta \alpha \biggr )^{2} 
+ \biggl( 
\frac{\partial R}{\partial d_{A}}\Delta d_{A} \biggr )^{2} \biggr ]^{1/2}.
\end{equation}
The angle uncertainty $\Delta \alpha $ was neglected by S12, so
$\Delta \alpha \sim 0$. Regarding the angular diameter distance, the small
uncertainties in the parameters of the cosmological models are such that
once propagated they change very little the value of $d_{A}$. So, we effectively
have $\Delta d_{A} \sim 0$.

\subsubsection{$\Delta \log (R_{eff})$}

Similarly to $\Delta \mu_{eff}$, the uncertainty $\Delta \log
(R_{eff})$ is derived from the linear fit of the Virgo galaxy cluster, 
\begin{equation}
\log (R_{eff}) = A_{\log (R_{eff})}n + B_{\log (R_{eff})}, 
\end{equation}
where $A$ and $B$ are the parameters given by equation
(\ref{eq.KormendyReffN}). Therefore,
\begin{eqnarray}
\Delta \log (R_{eff}) &=& \biggl \{ \biggl[ 
\frac{\partial \log (R_{eff})}{\partial A_{\log (R_{eff})}}\Delta A_{\log (R_{eff})} 
\biggr ]^{2} + \biggl[
\frac{\partial \log (R_{eff})}{\partial n}\Delta n \biggr ]^{2} \nonumber \\ 
 && + \biggl[ 
\frac{\partial \log (R_{eff})}{\partial B_{\log (R_{eff})}}\Delta B_{\log (R_{eff})} 
\biggr ]^{2} \biggr \}^{1/2},
\end{eqnarray}
which may be rewritten as,
\begin{equation}
\Delta \log (R_{eff}) = 
\biggl \{ \bigl [n\Delta A_{\log (R_{eff})} \bigr ]^2 + (A\Delta n)^2 + 
\bigl [\Delta B_{\log (R_{eff})} \bigr ]^2 \biggr \}^{1/2}. 
\end{equation}
Just like in $\Delta \mu_{eff}$, $\Delta A_{\log (R_{eff})}$
and $\Delta B_{\log (R_{eff})}$ come from the linear fit and $\Delta n$
is given by observations.

\subsubsection{$\Delta \mu_{re,\nu_{re}}$} 

Putting together all these expressions for uncertainties in
equation (\ref{erromu}) and remembering the relationship between
$b_n$ and $n$ (Eq.\ \ref{ebn4}), the uncertainty in the theoretical
prediction of the surface brightness yields,
\begin{eqnarray}
\Delta \mu_{re, \nu_{re}} &=& 
\Biggl ( (n\Delta A_{\mu_{eff}})^2+(A_{\mu_{eff}} \Delta n)^2 
+ (\Delta B_{\mu_{eff}})^2 + \nonumber \\
&& + \biggl [\biggl ( \frac{2.5}{\ln 10}\frac{{\mathrm d}b_n}
{{\mathrm d}n}\biggl \{\bigg[ \frac{R}{10^{\log (R_{eff})}}\biggr]^{1/n}
-1\biggr \} - \nonumber \\
&& - \frac{2.5b_n}{n^2\ln 10}\biggl [\frac{R}{10^{\log (R_{eff})}}
\biggr]^{1/n} \log \biggl \{ \frac{R}{10^{\log (R_{eff})}} \biggr \}
\biggr)\Delta n\biggr ]^2 + \nonumber \\ 
&& + \biggl \{ \frac{2.5b_n}{n}\biggl [\frac{R}{10^{\log (R_{eff})}}
\biggr]^{1/n} \biggr \}^2 \biggl \{[n\Delta A_{\log (R_{eff})}]^2 + 
\nonumber \\ 
&& + [A_{\log (R_{eff})} \Delta n]^2 + [\Delta B_{\log (R_{eff})}]^2
\biggr \}\Biggr )^{1/2}.
\label{eq.errormhuTH}
\end{eqnarray}

\subsubsection{Comparing theory and observation}

Before we can actually compare our S12 galaxy subsample with
the theoretical predictions of the surface brightness profile, we still
need to estimate the spectral energy distribution (SED) $J$. We proceed
on this point from the very simple working assumption of a constant
value for the SED, since using the overall energy distribution of the
galaxies in the Virgo cluster in different bandwidths is beyond the
aims of this paper. Henceforth, we assume the working value of $J=0.5$. 

The scaling relations obtained from the galaxies of the Virgo
cluster, Eqs.\ (\ref{eq.KormendyMhuN}) and (\ref{eq.KormendyReffN}),
allow us to calculate $\mu_{eff}$ and $R_{eff}$ for a given S\'{e}rsic
index value. So, these two equations may be rewritten as,
\begin{equation}
\mu_{eff,V} = 0.38n +20.5,
\label{eq.muSR} 
\end{equation}
where its error is given by,
\begin{equation}
 \Delta \mu_{eff,V} = 0.09n +0.5,
\label{eq.emuSR}
\end{equation}
and 
\begin{equation}
\log R_{eff,V} = 0.21n -0.5,  
\label{eq.ReffSR}
\end{equation}
whose uncertainty is, 
\begin{equation}
\Delta \log R_{eff,V} = 0.03n + 0.1. 
\label{eq.eReffSR}
\end{equation}

\begin{figure*}
\includegraphics[width=8.8cm]{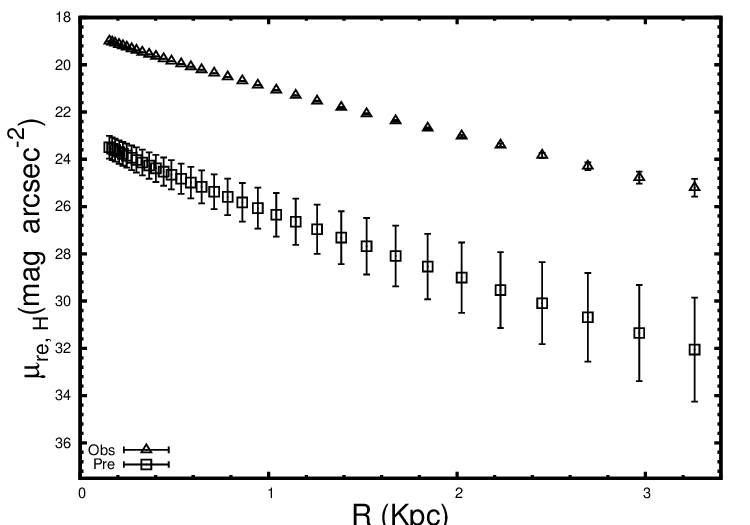}
\includegraphics[width=8.8cm]{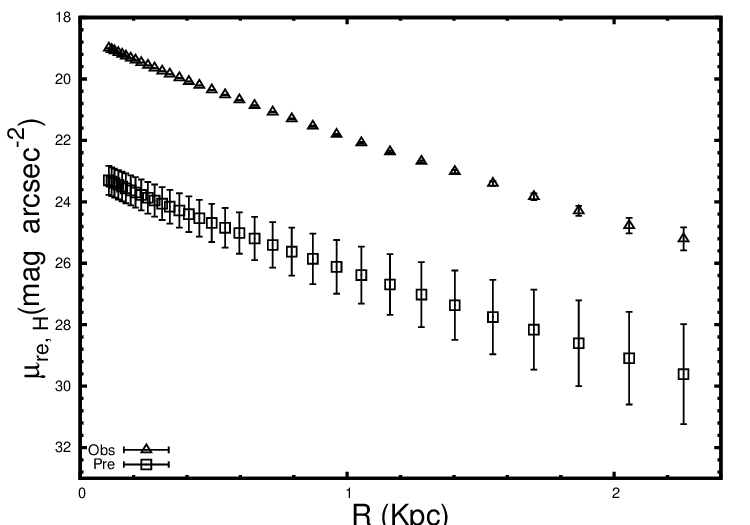}
\caption{\textit{Both}: galaxy ID 2.856 from S12 labeled as ``Obs'' at
$z = 1.759$ with $n = 1.2 \pm 0.08$. Theoretical prediction using Eqs.\
(\ref{eq.mhuTH}) and (\ref{eq.errormhuTH}) assuming the scaling relations
of Virgo cluster galaxies, labeled as ``Pre''.  \textit{Left}: Considering
$\Lambda$CDM cosmological model. \textit{Right}: Considering Einstein-de
Sitter cosmological model.}  
\label{figura1N2.856}
\end{figure*}

\begin{figure*}
\includegraphics[width=8.8cm]{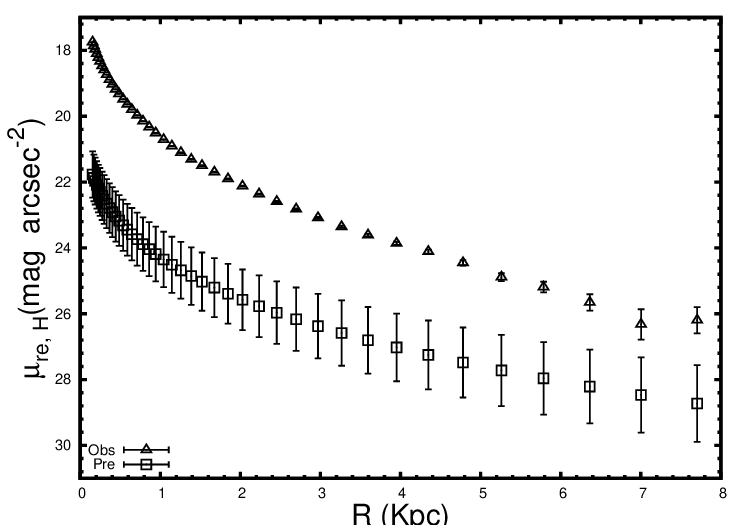}
\includegraphics[width=8.8cm]{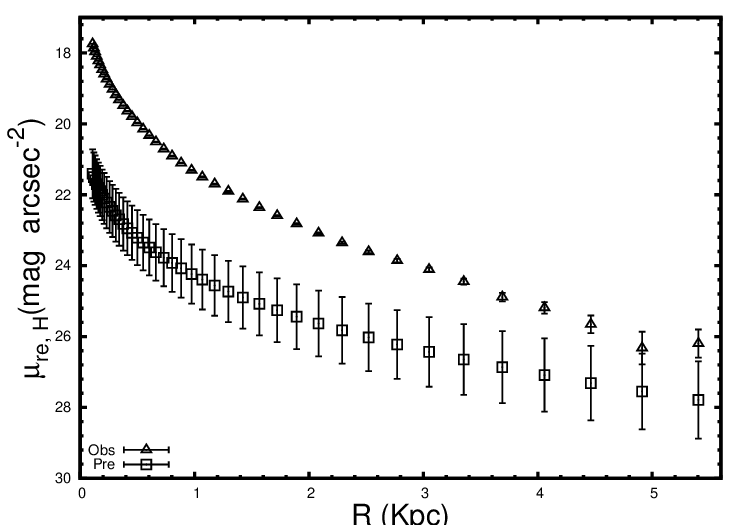}
\caption{\textit{Both}: galaxy ID 2.531 from S12 labeled as ``Obs'' at
$z = 1.598$ with $n = 4.08 \pm 0.3$. Theoretical prediction using Eqs.\
(\ref{eq.mhuTH}) and (\ref{eq.errormhuTH}) assuming the scaling relations
of Virgo cluster galaxies, labeled as ``Pre''. \textit{Left}: Considering
$\Lambda$CDM cosmological model. \textit{Right}: Considering Einstein-de
Sitter cosmological model.}  
\label{figura4N2.531}
\end{figure*}

\begin{figure*}
\includegraphics[width=8.8cm]{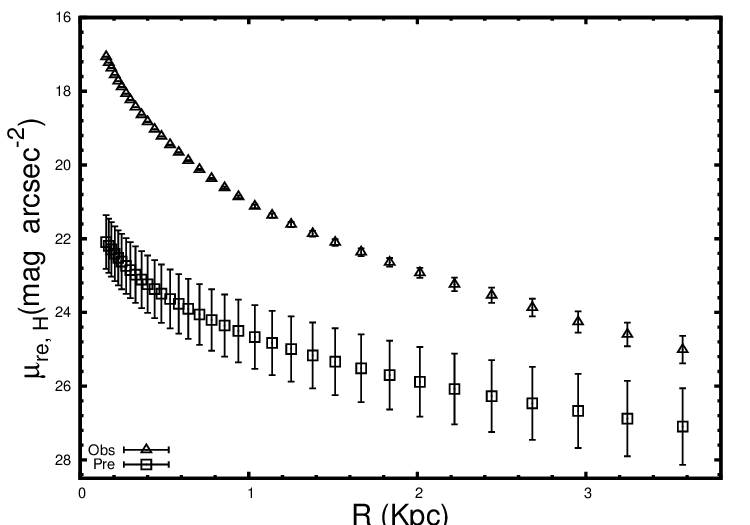}
\includegraphics[width=8.8cm]{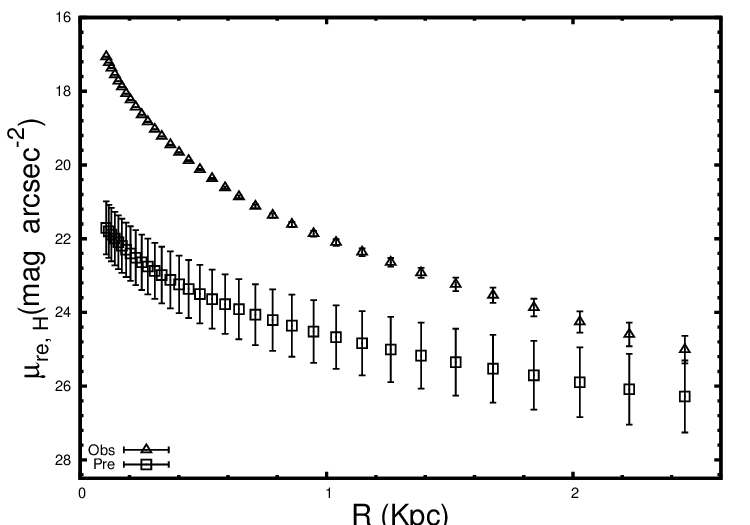}
\caption{\textit{Both}: galaxy ID 3.242 from S12 labeled 
as ``Obs'' at $z = 2.47$ with $n = 4.17 \pm 0.45$. 
Theoretical prediction using Eqs.\ (\ref{eq.mhuTH}) and (\ref{eq.errormhuTH}) 
assuming the scaling relations of Virgo cluster galaxies, 
labeled as ``Pre''. 
\textit{Left}: Considering $\Lambda$CDM cosmological model.  
\textit{Right}: Considering Einstein-de Sitter cosmological model.}  
\label{figura4N3.242}
\end{figure*}

\begin{figure*}
\includegraphics[width=8.8cm]{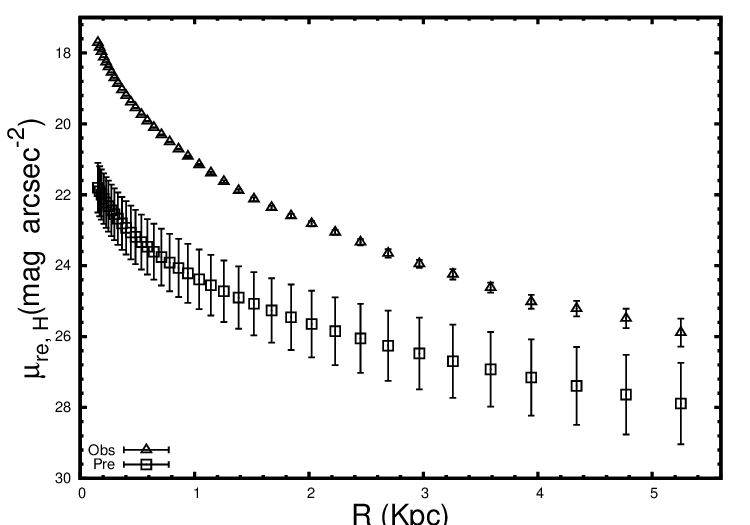}
\includegraphics[width=8.8cm]{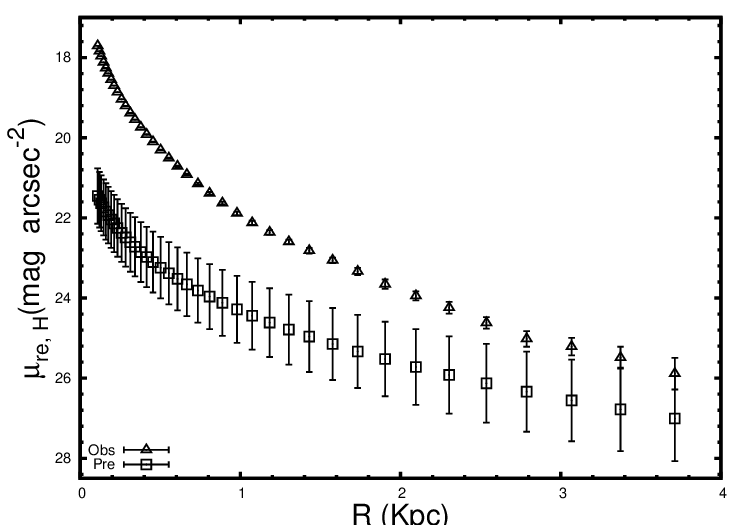}
\caption{\textit{Both}: galaxy ID 3.548 from S12 labeled 
as ``Obs'' at $z = 1.5$ with $n = 3.75 \pm 0.48$. 
Theoretical prediction using Eqs.\ (\ref{eq.mhuTH}) and (\ref{eq.errormhuTH}) 
assuming the scaling relations of Virgo cluster galaxies, 
labeled as ``Pre''. 
\textit{Left}: Considering $\Lambda$CDM cosmological model.  
\textit{Right}: Considering Einstein-de Sitter cosmological model.}  
\label{figura4N3.548}
\end{figure*}

\begin{figure*}
\includegraphics[width=8.8cm]{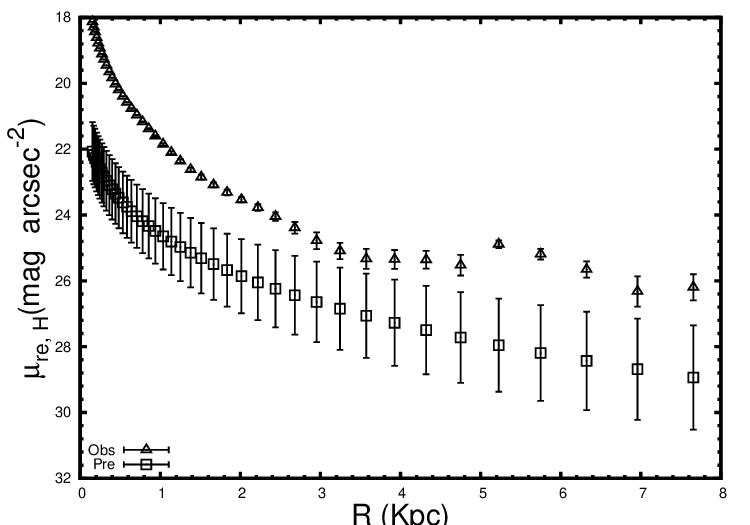}
\includegraphics[width=8.8cm]{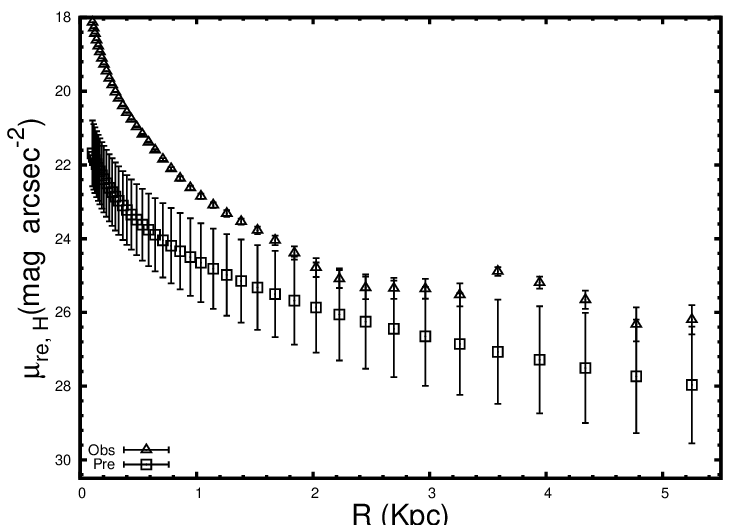}
\caption{\textit{Both}: galaxy ID 3.829 from S12 labeled 
as ``Obs'' at $z = 1.924$ with $n = 4.24 \pm 1.15$. 
Theoretical prediction using Eqs.\ (\ref{eq.mhuTH}) and (\ref{eq.errormhuTH}) 
assuming the scaling relations of Virgo cluster galaxies, 
labeled as ``Pre''. 
\textit{Left}: Considering $\Lambda$CDM cosmological model.  
\textit{Right}: Considering Einstein-de Sitter cosmological model.}  
\label{figura4N3.829}
\end{figure*}

\begin{figure*}
\includegraphics[width=8.8cm]{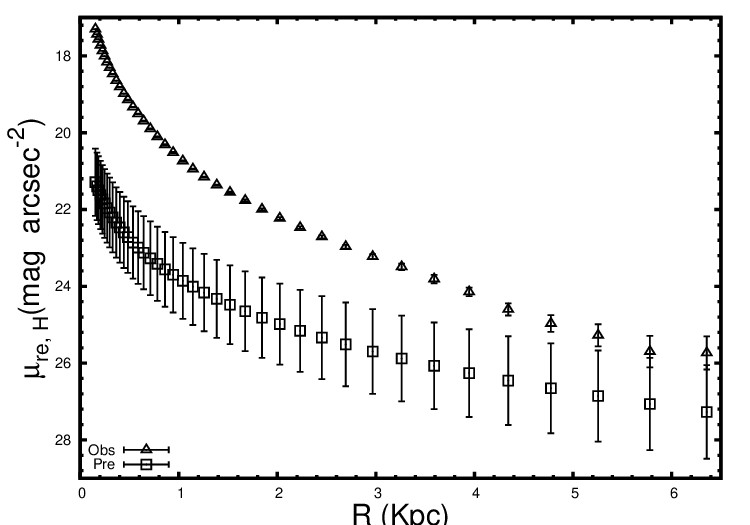}
\includegraphics[width=8.8cm]{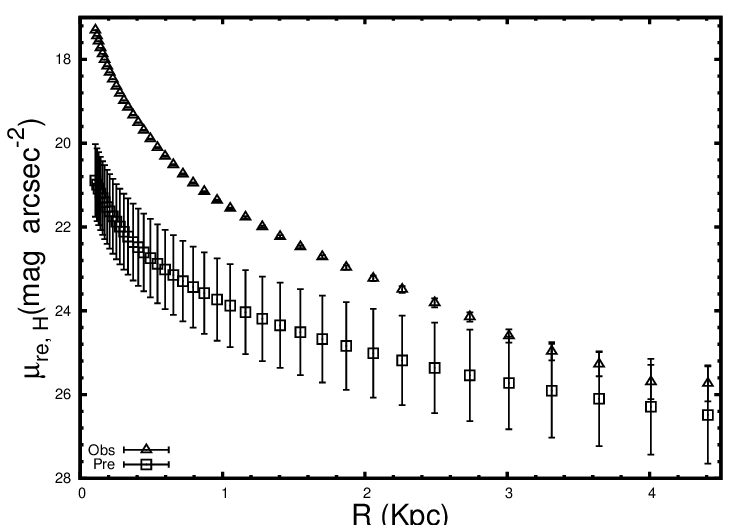}
\caption{\textit{Both}: galaxy ID 1.088 from S12 labeled 
as ``Obs'' at $z = 1.752$ with $n = 5.5 \pm 0.67$. 
Theoretical prediction using Eqs.\ (\ref{eq.mhuTH}) and (\ref{eq.errormhuTH}) 
assuming the scaling relations of Virgo cluster galaxies, 
labeled as ``Pre'' points. 
\textit{Left}: Considering $\Lambda$CDM cosmological model.  
\textit{Right}: Considering Einstein-de Sitter cosmological model.}  
\label{figura5N1.088}
\end{figure*}

\begin{figure*}
\includegraphics[width=8.8cm]{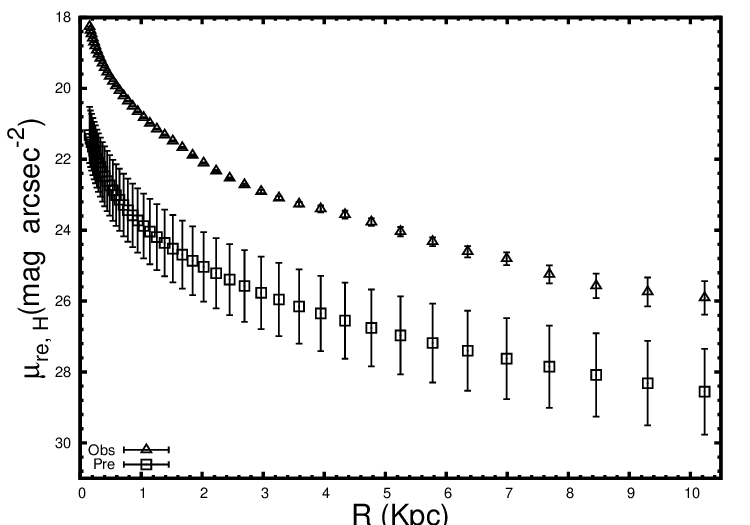}
\includegraphics[width=8.8cm]{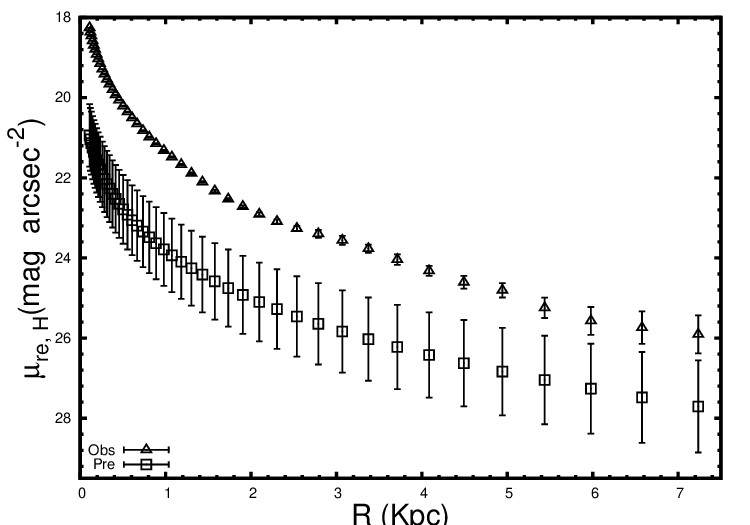}
\caption{\textit{Both}: galaxy ID 1.971 from S12 labeled 
as ``Obs'' at $z = 1.608$ with $n = 5.07 \pm 0.31$. 
Theoretical prediction using Eqs.\ (\ref{eq.mhuTH}) and (\ref{eq.errormhuTH}) 
assuming the scaling relations of Virgo cluster galaxies, 
labeled as ``Pre''. 
\textit{Left}: Considering $\Lambda$CDM cosmological model.  
\textit{Right}: Considering Einstein-de Sitter cosmological model.}  
\label{figura5N1.971}
\end{figure*}

\begin{figure*}
\includegraphics[width=8.8cm]{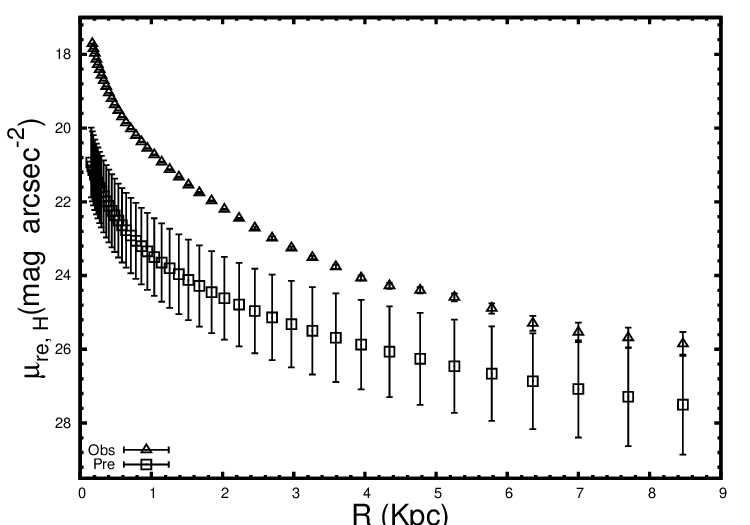}
\includegraphics[width=8.8cm]{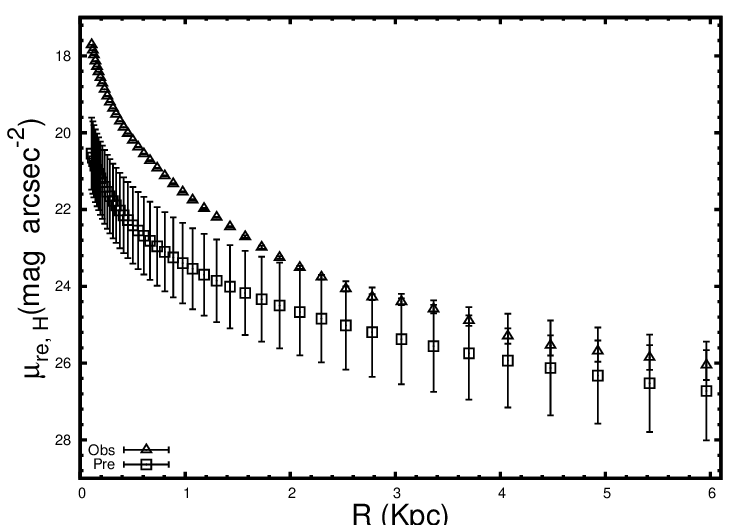}
\caption{\textit{Both}: galaxy ID 2.514 from S12 labeled 
as ``Obs'' at $z = 1.548$ with $n = 5.73 \pm 0.93$. 
Theoretical prediction using Eqs.\ (\ref{eq.mhuTH}) and (\ref{eq.errormhuTH}) 
assuming the scaling relations of Virgo cluster galaxies, 
labeled as ``Pre''. 
\textit{Left}: Considering $\Lambda$CDM cosmological model.  
\textit{Right}: Considering Einstein-de Sitter cosmological model.}  
\label{figura5N2.514}
\end{figure*}

\begin{figure*}
\includegraphics[width=8.8cm]{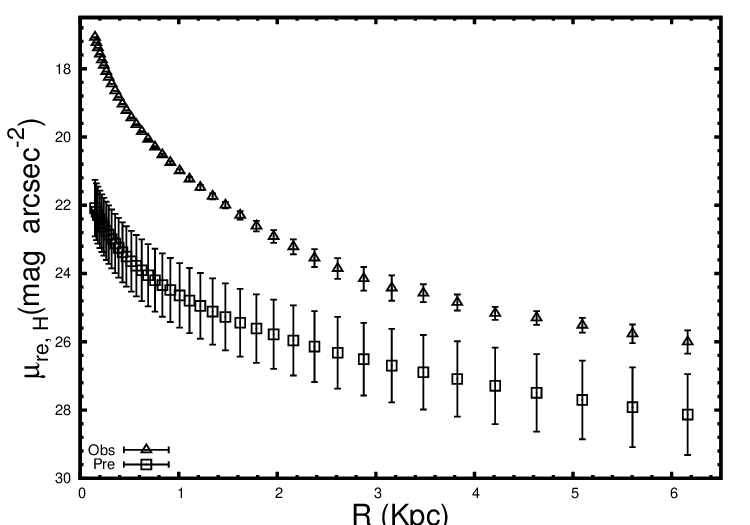}
\includegraphics[width=8.8cm]{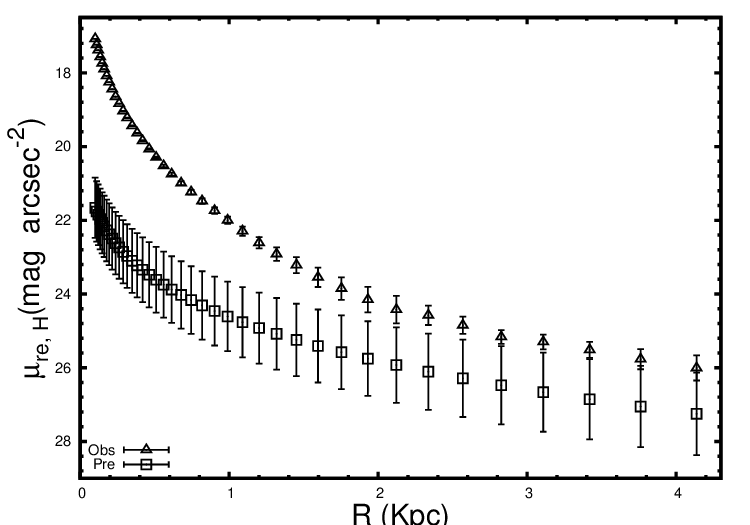}
\caption{\textit{Both}: galaxy ID 3.119 from S12 labeled 
as ``Obs'' at $z = 2.349$ with $n = 5.09 \pm 0.6$. 
Theoretical prediction using Eqs.\ (\ref{eq.mhuTH}) and (\ref{eq.errormhuTH}) 
assuming the scaling relations of Virgo cluster galaxies, 
labeled as ``Pre''. 
\textit{Left}: Considering $\Lambda$CDM cosmological model.  
\textit{Right}: Considering Einstein-de Sitter cosmological model.}  
\label{figura5N3.119}
\end{figure*}

\begin{figure*}
\includegraphics[width=8.8cm]{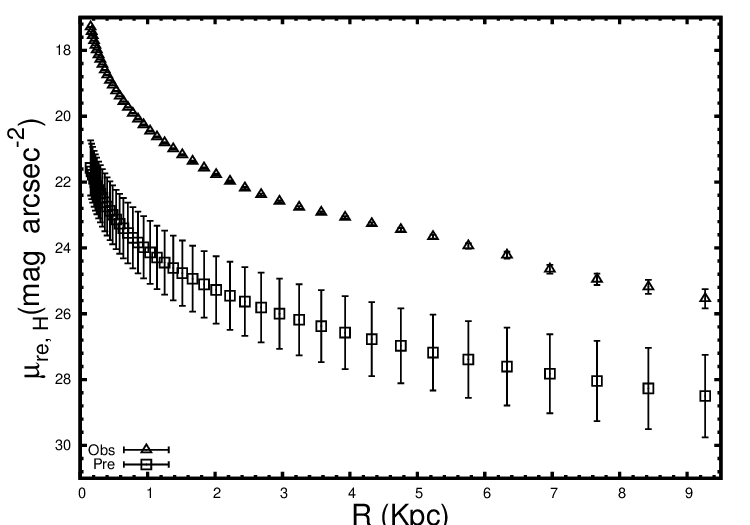}
\includegraphics[width=8.8cm]{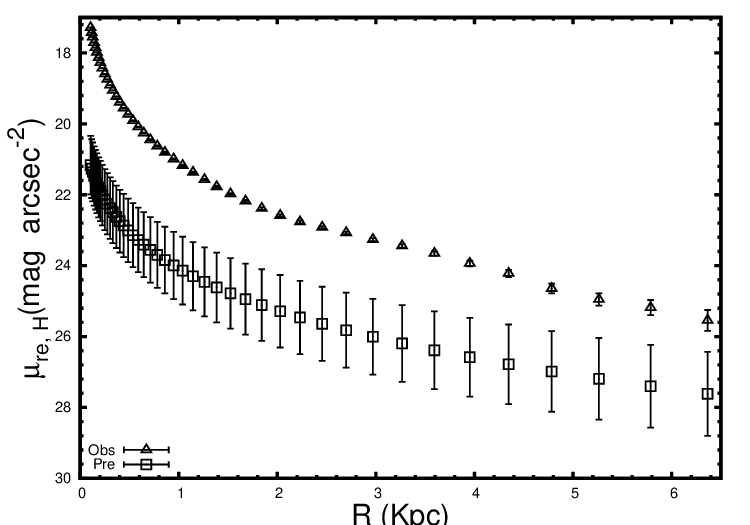}
\caption{\textit{Both}: galaxy ID 6.097 from S12 labeled 
as ``Obs'' at $z = 1.903$ with $n = 5.26 \pm 0.56$. 
Theoretical prediction using Eqs.\ (\ref{eq.mhuTH}) and (\ref{eq.errormhuTH}) 
assuming the scaling relations of Virgo cluster galaxies, 
labeled as ``Pre''. 
\textit{Left}: Considering $\Lambda$CDM cosmological model.  
\textit{Right}: Considering Einstein-de Sitter cosmological model.}  
\label{figura5N6.097}
\end{figure*}

Graphs showing the S12 data, labeled as ``Obs'', and the
theoretical predictions of the surface brightness profile, labeled as
``Pre'', in each of the two cosmological models adopted in this work,
$\Lambda$CDM and Einstein-de Sitter, are shown in Figs.\ \ref{figura1N2.856}
to \ref{figura5N6.097}. One can clearly see that the observed and predicted
results are very different, a result which shows that the Virgo cluster
galaxies do not behave as our S12 subsample. So, for our high redshift
galaxies to evolve into the Virgo cluster ones the scaling relations
parameters have to change in order to reflect such an evolution. One can
also see that the difference in the results when employing each of the
two cosmological models is not at all significant. Thus, our next step
is to work out the changes in the parameter to find out if the evolution
these galaxies have to sustain is strongly dependent on the assumed
underlying cosmology.

\section{Evolution to the Virgo cluster scaling parameters in $\Lambda$CDM and 
EdS cosmologies}\label{section7}    

As seen above, the theoretical prediction of the surface brightness
profiles obtained through scaling relations derived from data of the Virgo
galactic cluster are very different from the observed surface brightness of
S12 galaxies in both cosmological models studied here. Hence, in order to
ascertain the possible evolution that these galaxies would have to experience 
such that they end up with surface brightness equal to the ones in the Virgo
cluster, we need to look carefully at the parameter evolution of the scaling
relations. Assuming that they do not evolve through the S\'{e}rsic index $n$
(see above), we can study such an evolution via the effective brightness
and the effective radius.

Let $\mu_{eff,evo}$ and $\log R_{eff,evo}$ be, respectively, the
evolution of the effective brightness and the effective radius. We can
estimate these two quantities similarly to our previous calculation of
the scaling relations. For values of the S\'ersic index of our S12
galactic subsample, we can obtain their respective Virgo cluster galaxies
effective brightness and effective radius $\mu_{eff,V}$ and $\log R_{eff,V}$
by means of the expressions (\ref{eq.muSR}) to (\ref{eq.eReffSR}). We then
modify the linear fit parameters $A$ and $B$ in these expressions ($y=Ax+B$),
adjusting them so that the results approximate the points of the
theoretical predictions with S12's observed values in order to find
$\mu_{eff,Szo}$ and $\log R_{eff,Szo}$, that is, the values of effective
brightness and effective radius of our subsample of S12's galaxies.
Therefore, both $\mu_{eff,evo}$ and $\log R_{eff,evo}$ are given as
follows,
\begin{equation}
 \mu_{eff,evo} = \mu_{eff,Szo} - \mu_{eff,V},
\label{eq.MUevo}
\end{equation}
\begin{equation}
\log R_{eff, evo} = \log R_{eff, Szo} - \log R_{eff, V}.
\label{eq.Reffevo}
\end{equation}
The respective uncertainties yield,
\begin{equation}
\Delta \mu_{eff, evo} = \Delta \mu_{eff, Szo} + \Delta \mu_{eff, V},
\label{eq.eMUevo}
\end{equation}
\begin{equation}
\Delta \log R_{eff, evo} = \Delta \log R_{eff, Szo} +
\Delta \log R_{eff, V}. 
\label{eq.eReffevo}
\end{equation}

Table \ref{table2} shows the effective brightness and effective
radius for the Virgo cluster galaxies obtained with S\'ersic indexes
$n$ equal to those in our S12 subsample presented in Table \ref{table1}.
These quantities were obtained using the Virgo scaling relations.
\begin{table}
\caption[]
{Effective brightness $\mu_{eff, V}$ and effective radius
$\log R_{eff, V}$ obtained with Virgo cluster galaxies scaling
relations for values of $n$ equal to those in our subsample of
S12 galaxies.}
\begin{center}
   \begin{tabular}{@{\extracolsep{\fill}}l c r 
		   @{\extracolsep{\fill}}} \hline \hline
		  ID & $\mu_{eff, V}$ & $\log R_{eff, V}$ \\
      \hline
2.856 & 21.0 $\pm$ 0.7 & -0.3 $\pm$ 0.2\\
3.548 & 21.9 $\pm$ 0.9 & 0.3 $\pm$ 0.3 \\
2.531 & 22.1 $\pm$ 0.9 & 0.4 $\pm$ 0.3 \\
3.242 & 22.1 $\pm$ 0.9 & 0.4 $\pm$ 0.3 \\
3.829 & 22.1 $\pm$ 0.9 & 0.4 $\pm$ 0.3 \\
1.971 & 22 $\pm$ 1     & 0.6 $\pm$ 0.3 \\
3.119 & 22 $\pm$ 1     & 0.6 $\pm$ 0.3 \\
6.097 & 22 $\pm$ 1     & 0.6 $\pm$ 0.3 \\
1.088 & 23 $\pm$ 1     & 0.6 $\pm$ 0.3 \\
\label{table2}
\end{tabular}
\end{center}
\end{table}
Table \ref{table3} shows the same quantities for the adjusted 
parameters of our subsample of S12 galaxies using the two cosmological
models considered here and Table \ref{table4} presents the
evolution of the effective brightness and effective radius calculated
using eqs.\ (\ref{eq.MUevo}) to (\ref{eq.eReffevo}) in both cosmological
models in the V band.
\begin{table}
\caption[]
{Effective brightness $\mu_{eff, Szo}$ and effective radius
$\log R_{eff, Szo}$ of the adjusted parameters from the galaxies 
selected from S12 for values of $n$ equal to
the ID galaxies in the V band in the two cosmological models
considered in this paper.} 
\begin{center}
  \small \begin{tabular}{@{\extracolsep{\fill}}l c c c r 
		   @{\extracolsep{\fill}}} \hline \hline
		  ID & $\mu_{eff, Szo, \Lambda CDM}$ & 
$\log R_{eff, Szo, \Lambda CDM}$ & $\mu_{eff, Szo, EdS}$ & 
$\log R_{eff, Szo, EdS}$ \\
      \hline
2.856 & 16.5 $\pm$ 0.6 & -0.1 $\pm$ 0.1 & 16.7 $\pm$ 0.6 & -0.1 $\pm$ 0.1\\
3.548 & 16.5 $\pm$ 0.8 & -0.2 $\pm$ 0.3 & 16.4 $\pm$ 0.8 & -0.3 $\pm$ 0.2\\
2.531 & 16.7 $\pm$ 0.9 &  0.0 $\pm$ 0.3 & 17.1 $\pm$ 0.9 & -0.1 $\pm$ 0.3\\
3.242 & 14.9 $\pm$ 0.9 & -0.4 $\pm$ 0.3 & 14.3 $\pm$ 0.9 & -0.6 $\pm$ 0.3\\
3.829 & 18.1 $\pm$ 0.9 & 0.2 $\pm$ 0.2  & 17.1 $\pm$ 0.9 & -0.1 $\pm$ 0.3\\
1.971 & 19.3 $\pm$ 0.9 & 0.5 $\pm$ 0.3  & 20 $\pm$ 1     &  0.5 $\pm$ 0.3\\
3.119 & 16 $\pm$ 1     & -0.1 $\pm$ 0.3 & 16 $\pm$ 1     & -0.3 $\pm$ 0.3\\
6.097 & 17 $\pm$ 1     & 0.3 $\pm$ 0.3  & 18 $\pm$ 1     &  0.3 $\pm$ 0.3\\
1.088 & 16 $\pm$ 1     & 17 $\pm$ 1     & 17 $\pm$ 1     & -0.2 $\pm$ 0.3\\

\label{table3}
\end{tabular}
\end{center}
\end{table}
\begin{table}
\caption[]
{Evolution of the effective brightness $\mu_{eff, evo}$ and effective
radius $\log R_{eff, evo}$ of the galaxies selected from S12 
for values of $n$ equal to the ID galaxies in the V band in the two
cosmological models considered here.} 
\begin{center}
   \begin{tabular}{@{\extracolsep{\fill}}l c c c r 
		   @{\extracolsep{\fill}}} \hline \hline
		  ID & $\mu_{eff, evo, \Lambda CDM}$ 
& $\mu_{eff, evo, EdS}$ & $\log R_{eff, evo, \Lambda CDM}$ 
& $\log R_{eff, evo, EdS}$ \\
      \hline
2.856 & -5 $\pm$ 2 & -4 $\pm$ 2 & 0.2 $\pm$ 0.3  & 0.2 $\pm$  0.3\\
3.548 & -6 $\pm$ 2 & -6 $\pm$ 2 & -0.4 $\pm$ 0.4 & -0.6 $\pm$ 0.5\\
2.531 & -5 $\pm$ 2 & -5 $\pm$ 2 & -0.4 $\pm$ 0.5 & -0.5 $\pm$ 0.5\\
3.242 & -7 $\pm$ 2 & -7 $\pm$ 2 & -0.7 $\pm$ 0.5 & -1.0 $\pm$ 0.4\\
3.829 & -4 $\pm$ 2 & -5 $\pm$ 2 & -0.2 $\pm$ 0.3 & -0.5 $\pm$ 0.5\\
1.971 & -3 $\pm$ 2 & -2 $\pm$ 2 &  0.0 $\pm$ 0.5 &  0.0 $\pm$ 0.5\\
3.119 & -6 $\pm$ 2 & -6 $\pm$ 2 & -0.7 $\pm$ 0.5 & -0.8 $\pm$ 0.5\\
6.097 & -5 $\pm$ 2 & -4 $\pm$ 2 & -0.3 $\pm$ 0.5 & -0.2 $\pm$ 0.5\\
1.088 & -6 $\pm$ 2 & -6 $\pm$ 2 & -0.7 $\pm$ 0.6 & -0.8 $\pm$ 0.6\\

\label{table4}
\end{tabular}
\end{center}
\end{table}

The results show that the evolution that will have to occur so that
the S12 high redshift galaxies have effective brightness and effective
radius equal to the ones in the Virgo cluster is similar in both cosmological
models. Specifically, it seems that in EdS model the effective radius
evolution is higher than the one occurred in the $\Lambda$CDM model.
Nevertheless, the evolution of the effective surface brightness is almost
the same in both models. We also note that the difference in the S\'{e}rsic
index values do not appear to affect our results, which also clearly show
that the uncertainties in the measurements of both quantities we deal with
here are just too high to allow us to distinguish the underlying
cosmological model that best represents the data. Basically our methodology
is limited by the uncertainties, at least as far as the S12 subset data is
concerned.

As final words, this work is based on the assumption that the galaxies
we compare belong to a group whose members share at least one common
feature regardless of the redshift, otherwise comparison among them becomes
impossible. In other words, our basic assumption is that our selected
galaxies belong to a homogeneous class of galaxies. In the analysis we
carried out above we grouped our galaxies using only the S\'{e}rsic
indexes, this therefore being our defining criterion of a homogeneous
class of objects. So, in a sense we followed Ellis \& Perry (1979) and
adopted morphology as our definition of a homogeneous class. 
  
\section{Summary and conclusions}
\label{section8}

In this paper we have compared high redshift surface brightness
observational data with theoretical surface brightness predictions for
two cosmological models, namely the $\Lambda$CDM and Einstein-de Sitter,
in order to test if such comparison allows us to distinguish the
cosmology that best fits the observational data. 
We started by reviewing the expressions for the emitted and received
bolometric source brightness and then obtained their respective specific
expressions in the context of galactic surface brightness (Ellis 1971;
Ellis \& Perry 1979).

Using the S\'{e}rsic profile, we have obtained scaling relations 
between the surface effective brightness $\mu_{eff}$ and S\'{e}rsic index,
as well as between the effective radius $R_{eff}$ and $n$, for the Virgo
cluster galaxies using Kormendy et al.\ (2009) data. Assuming this scaling
relation, we have calculated theoretical predictions of the surface
brightness and compared them with some of the observed surface brightness
profiles of high-redshift galaxies in a subsample of Szomoru et al.\ (2012)
galaxies in the two cosmological models considered here. Our results showed
that although the S\'{e}rsic profile fits well the observed brightness,
the results for surface brightness is different from the theoretical
predictions. Such difference was used to calculate the amount of evolution
that the high redshift galaxies would have to experience in order to
achieve the Virgo cluster structure once they arrive at $z \sim 0$. We
concluded from our results that the cosmological evolution is quite
similar in the two models considered in this paper. We also noted that
galaxies having different S\'{e}rsic indexes do not seem to follow a
different evolutionary path.

Overall, with the data and errors available for the chosen subset of
galactic profiles used here we cannot distinguish between the two
different cosmological models assumed in this work.  That is, assuming
that the high redshift galaxies will evolve to have features similar
to the ones found in the Virgo cluster, it is not possible to conclude
which cosmological model will predict theoretical surface brightness
curves similar to the observed ones due to the high uncertainties
in the data used here. We also noted that the S\'{e}rsic index does not
seem to play any significant evolutionary role, as the evolution we
discussed is apparently not affected by the value of $n$. Nevertheless,
this work used only the S\'ersic index to define a homogeneous class of
objects. 

Considering that the results shown above depend on the chosen
underlying cosmology it is reasonable to ask if the use of a different
enough cosmological model will produce relatively large effects in the
quantities analyzed here. On this regard, the most different cosmologies
still capable of describing observed features of the Universe are the
inhomogenous cosmological models (Bolejko 2006; Bolejko et al.\ 2011;
Krasi\'nski 2011; Hellaby and Walters 2018, and references therein).
Those, however, come with much greater degrees of freedom as compared
to the standard cosmology in the form of arbitrary functions which must
be defined a priori in terms of some kind of presumably observable
features, particularly mass distribution (Bolekjo et al.\ 2011, \S 4),
which can be defined in such a way that the resulting model becomes an
almost standard cosmology as far as some observables are concerned.
This is the case of the simplest inhomogeneous model, the
Lema\^itre-Tolman-Bondi (LTB) cosmology, whose arbitrary functions can
be established in such a way that in some cases the final cosmology
differs little in terms of some chosen observational features from the
results obtained with the standard cosmology (e.g., Ribeiro 1993,
Iribarrem et al.\ 2014, Lopes et al.\ 2017). If, however, one goes to
more extreme inhomogeneous cosmologies specifications their high
nonlinearity prevents us from reaching any conclusion beforehand. One
has to actually carry out the calculations in order to see if there is
any relative large effects in the quantities analyzed here.

The main point is that the intrinsic high nonlinearity of
relativistic cosmological models may lead to different results for each
observational quantity that is calculated. For instance, a particular
LTB model that provides observational results similar to the standard model
for a certain set of observations might end up producing entirely different
ones for other set of observations. In addition, since it seems that no
inhomogeneous model has so far been used to describe the problem discussed
here, if and how much inhomogeneous cosmologies, LTB or otherwise, would
affect the galactic surface brightness profiles in the form studied in this
paper remains an open problem. Therefore, it seems that to start looking at
this problem one has to choose a simple LTB model already used by other
authors, which describes some basic observed features such as voids, and
then carry out the calculations in order to draw comparisons.

Summing up those results, it seems reasonable that future studies
of this kind should also select galaxies based on other features
besides morphology in order to increase the number of common properties
between high and low redshift galaxies, and, of course, using different
data samples than those adopted here. More common features such as the
S\'ersic index are essential for a better definition of a homogeneous
class of cosmological objects whose observational features are possibly
able to distinguish among different cosmological scenarios. However,
care should be taken to avoid features which can possibly suffer
dramatic cosmological evolution.

\section*{Acknowledgments}
We are grateful to a referee for useful suggestions. I.O.S.\ thanks
the financial suport from \textit{Coordena\c{c}\~{a}o de
Aperfei\c{c}oamento de Pessoal de N\'{\i}vel Superior - Brasil (CAPES) -}
Finance Code 001.

\section*{Data Availability}
All data generated or analysed during this study are included in this
published article.

\section*{Competing Interests}
The authors have no competing interests to declare that are relevant to
the content of this article.

\label{lastpage}

\begin{thebibliography}{1}
\bibitem[Bolejko (2006)]{Bolejko2006}
	{Bolejko, K.} 2006, Proc.\ 13th Young Scientists' Conference on
	Astron.\ \&  Space Phys., Kyiv, Ukraine; A.\ Golovin, G.\
	Ivashchenko \& A.\ Simon (eds), arXiv:astro-ph/0607130 

\bibitem[Bolejko et al. (2011)]{Bolejko2011}
	{Bolejko, K., C\'el\'erier, M. N., \& Krasi\'nski, A.} 2011, Class.
        Quantum Grav., 28, 164002, arXiv:1102.1449

\bibitem[Bradt (2004)]{Bradt04}
  {Bradt, H.} 2004, Astronomy Methods: A Physical Approach to Astronomical
  Observations, UK: Cambridge University Press, 2004

\bibitem[Caon, Capaccioli \& D'Onofrio (1993)]{Caon93}
  {Caon, N., Capacccioli, M., \& D'Onofrio, M.} 1993, 
  MNRAS, 265, 1013

\bibitem[Capaccioli (1989)]{Capaccioli89}
  {Capaccioli, M.} 1989, 
  in World of Galaxies, 
  eds. H.G. Corwin Jr.\ \& L.\ Bottinelli (Berlin: Springer-Verlag), 208

\bibitem[Chakrabarty \& Jackson (2009)]{Chakrabarty09}
  {Chakrabarty, D. \& Jackson, B.} 2009, 
  A\&A, 498, 615

\bibitem[Ciotti \& Bertin (1991)]{CiottiL91}
  {Ciotti, L.} 1991, 
  A\&A, 249, 99

\bibitem[Ciotti \& Bertin (1999)]{Ciotti91}
  {Ciotti, L. \& Bertin, G.} 1999, 
  A\&A, 352, 447

\bibitem[Coppola, La Barbera \& Capaccioli (2009)]{Coppola09}
  {Coppola, G., La Barbera, F., \& Capaccioli, M.} 2009, 
  PASP, 121, 437

\bibitem[Davies et al. 1988]{Davies88}
  {Davies, J. I., Phillipps, S., Cawson, M. G. M., Disney, M. J., et al.} 1988, 
  MNRAS, 232, 239

\bibitem[Djorgovski (1987)]{Djorgovski87}
  {Djorgovski, S. \& Davis, M.} 1987, 
  AJ, 313, 59

\bibitem[D'Onofrio (1994)]{D'nofrio94}
  {D'Onofrio, M., Capaccioli, M., \& Caon, N.} 1994, 
  MNRAS, 271, 523

\bibitem[Ellis (1971)]{Ellis71}
  {Ellis, G. F. R.} 1971, 
  General Relativity and Cosmology, Proc.\ of the International School of
  Physics ``Enrico Fermi'', R. K. Sachs, New York: Academic Press;
  reprinted in Gen.\ Rel.\ Grav., 41 (2009) 581

\bibitem[Ellis (2006)]{Ellis06}
  {Ellis, G. F. R.} 2006, Handbook in Philosophy of Physics, 
   Ed. J. Butterfield and J. Earman. Dordrecht: Elsevier, 1183; 
  arXiv:astro-ph/0602280

\bibitem{ellis07}{Ellis, G.\ F.\ R.\ 2007, Gen.\ Rel.\ Grav., 39,
        1047}

\bibitem[Ellis (1985)]{Ellis85}
  {Ellis, G. F. R., Nel, S. D., Maartens, R., Stoeger, W. R., et al.} 1985,
  Phys. Rep., 124, 315

\bibitem[Ellis (1979)]{Ellis79}
  {Ellis, G. F. R. \& Perry, J. J.} 1979, 
  MNRAS, 187, 357

\bibitem[Ellis (1984)]{Ellis84}
  {Ellis, G. F. R., Sievers, A. W., \& Perry, J. J.} 1984, 
  AJ, 89, 1124

\bibitem[Etherington (1933)]{Etherington33}
  {Etherington, I. M. H.} 1933, 
  Philosophical Magazine, 15, 761;  reprinted in Gen.\ Rel.\ Grav.\
  39 (2007) 1055

\bibitem[Faber (1976)]{Ellis76}
  {Faber, S. M. \& Jackson, R. E.} 1976, 
  AJ, 204, 668

\bibitem[Graham (2001)]{Graham01}
  {Graham, A. W.} 2001, 
  AJ, 121, 820

\bibitem[Graham (2002)]{Graham02}
  {Graham, A. W.} 2002, 
  MNRAS, 334, 859

\bibitem[Graham \& Driver (2005)]{Graham05}
  {Graham, A. W. \& Driver, S. P.} 2005, 
  Publications of the Astronomical Society of Australia,
  22, 118

\bibitem[Hellaby \& Walters (2018)]{Hellaby18}
  {Hellaby, C. \&  Walters, A.} 2018, Journal of Cosmology and
  Astroparticle Physics, (02) 015, arXiv:1708.01031

\bibitem[Iribarrem et al. (2014)]{Iribarrem14}
  {Iribarrem, A., Andreani, P., February, S., Gruppioni, C., Lopes, A. R.,
  Ribeiro, M. B. \& Stoeger, W. R.} 2014, Astron. Astrophys. 563, A20,
  arXiv:1401.6572

\bibitem[Komatsu et al. (2009)]{Komatsu09}
  {Komatsu, E., Dunkley, J., Nolta, M. R., Bennett, C.L., et al.} 2009, 
  ApJS, 180, 330 

\bibitem[Kormendy (1977)]{Kormendy77}
  {Kormendy, J.} 1977, 
  AJ, 218, 333

\bibitem[Kormendy et al. (2009)]{Kormendy09}
  {Kormendy, J., Fisher, D. B., Cornell, M. E., \& Bender, R.} 2009, 
  ApJS, 182, 216

\bibitem[Krasinski (2011)]{Krasinski2011}
  {Krasi\'nski, A.} 2011, Acta Physica Polonica B42, 2263, arXiv:1110.1828v2

\bibitem[Kristian \& Sachs (1966)]{Kristian66}
  {Kristian, J. \& Sachs, R. K.} 1966, 
  ApJ, 143, 379; 
  reprinted in Gen.\ Rel.\ Grav., 43, 337, 2011

\bibitem[La Barbera et al. (2005)]{LaBarbera05}
  {La Barbera, F., Covone, G., Busarello, G., Capaccioli, M., et al.} 2005, 
  MNRAS, 358, 1116

\bibitem[Laurikainen et al. (2010)]{Laurikainen10}
  {Laurikainen, E., Salo, H., Buta, R., Knapen, J. H., et al.} 2010,
  MNRAS, 405, 1089

\bibitem[Lopes et al. (2017)]{Lopes17}
{Lopes, A. R., Gruppioni, C. G., Ribeiro, M. B.,  Pozzetti, L., February,
  S., Ilbert, O., Pozzi, F.} 2017, MNRAS, 471, 3098 

\bibitem[Mazure \& Capelato (2002)]{Mazure02}
  {Mazure, A. \& Capelato, H. V.} 2002, 
  A\&A, 383, 384

\bibitem[Naab \& Trujillo]{Naab06}
  {Naab, T. \& Trujillo, I.} 2006, 
  MNRAS, 369, 625

\bibitem[Olivares-Salaverri \& Ribeiro (2009)]{Olivares-Salaverri09}
  {Olivares-Salaverri, I. \& Ribeiro, M. B.} 2009, 
  Memorie della Societ\`a Astronomica Italiana, 80, 925; arXiv:0911.3035

\bibitem[Olivares-Salaverri \& Ribeiro (2010)]{Olivares-Salaverri10}
  {Olivares-Salaverri, I. \& Ribeiro, M. B.} 2010, 
  Highlights of Astronomy, 15, 329

\bibitem[Prugniel \& Simien (1997)]{Prugniel97}
  {Prugniel, P. \& Simien, F.} 1997, 
  A\&A, 321, 111

\bibitem[Ribeiro (1993)]{Ribeiro03}
  {Ribeiro, M. B.} 1993,	
  Astrophys. J., 415, 469; arXiv:0807.1021

\bibitem[Ribeiro (2005)]{Ribeiro05}
  {Ribeiro, M. B.} 2005,	
  A\&A, 429, 65; arXiv:astro-ph/0408316

\bibitem[S\'{e}rsic (1968)]{Sersic68}
  {S\'{e}rsic, J. L.} 1968, Atlas de galaxias australes, 
  Observatorio Astron\'{o}mico, Cordoba

\bibitem[Szomoru (2012)]{Szomoru12}
  {Szomoru, D., Franx, M., \& Van Dokkum, P. G.} 2012, 
  ApJ, 749, 121 \textbf{(S12)}

\bibitem[Tully (1977)]{Tully77}
  {Tully, R. B., \& Fisher, J. R.} 1977, 
  A\&A, 54, 661

\bibitem[Trujillo (2001)]{Trujillo01}
  {Trujillo, I., Graham, A. W., \& Caon, N.} 2001, 
  MNRAS, 326, 869

\end{thebibliography}
\end{document}